\documentclass[%
reprint,
superscriptaddress,
nofootinbib,
 amsmath,amssymb,
 aps,
prl,
]{revtex4-1}

\usepackage{graphicx,color}
\usepackage{dcolumn}
\usepackage{bm}
\usepackage{hyperref}

\begin{document}


\title{Frozen state and spin liquid physics in Na$_4$Ir$_3$O$_8$: an NMR study}

\author{A.C. Shockley}
\email{acshoc@gmail.com}
\affiliation{Laboratoire de Physique des Solides, Universit\'{e} Paris-Sud 11, UMR CNRS 8502, 91405 Orsay, France}
\author{F. Bert}
\affiliation{Laboratoire de Physique des Solides, Universit\'{e} Paris-Sud 11, UMR CNRS 8502, 91405 Orsay, France}
\author{J-C. Orain}
\affiliation{Laboratoire de Physique des Solides, Universit\'{e} Paris-Sud 11, UMR CNRS 8502, 91405 Orsay, France}
\author{Y. Okamoto}
\affiliation{Department of Applied Physics, Nagoya University, Furo-cho, Chikusa-ku, Nagoya 464-8603, Japan}
\author{P. Mendels}
\affiliation{Laboratoire de Physique des Solides, Universit\'{e} Paris-Sud 11, UMR CNRS 8502, 91405 Orsay, France}

\date{\today}

\begin{abstract}
Na$_4$Ir$_3$O$_8$ is a unique case of a hyperkagome 3D corner sharing triangular lattice which can be decorated with quantum spins. It has spurred a lot of theoretical interest as a spin liquid candidate. We present a comprehensive set of NMR data taken on both the $^{23}$Na and $^{17}$O sites. We show that disordered magnetic freezing of all Ir sites sets in below $T_f \sim 7$~K, well below $J=300$~K, with a drastic slowing down of fluctuations to a static state revealed by our $T_1$ measurements. Above typically $2\;T_f$, physical properties are relevant to the spin liquid state induced by this exotic geometry. While the shift data shows that the susceptibility levels off below 80~K, 1/$T_1$ has little variation from 300~K to $2\;T_f$. We discuss the implication of our results in the context of published experimental and theoretical work.
\end{abstract}

\pacs{75.10.Kt, 76.60.-k,76.30.-v,76.75.+i}
\maketitle


Geometric frustration has long been known as an essential ingredient to stabilize a quantum spin liquid (QSL) state in more than one dimension (1D)\cite{balents2010naturereviewspinliq,misguichlhuillierbook}. Since experimental realizations are rare, Herbertsmithite \cite{shores2005jacsherbdisc} triggered a lot of excitement with its discovery in 2005. It is the most well known example of a two dimensional (2D) QSL where $S=\frac{1}2$ Cu$^{2+}$ spins form a perfect kagome lattice of {\it corner-sharing} triangles with dominant Heisenberg interactions, no sign of ordering \cite{helton2007prlherb,mendels2007prlquantmag}and a continuum of excitations interpreted as a fractionalization of excitations into $S=\frac{1}2$ spinons \cite{han2012natureherbneutron}, similar to 1D $S=\frac{1}2$ Heisenberg chains. Recently other routes to 2D QSL physics have been explored such as frustration induced by competing interactions with Kapellasite \cite{fak2012prlkapellasite,kermarrec2014prbkapellasite} and, outside of cuprates, the kagome vanadate DQVOF with $S=\frac{1}2$ V$^{4+}$  \cite{clark2013prlvof}, Mo$^{5+}$ double perovskite \cite{deVries2010prlmo}, and organic compounds which are driven by proximity to a Mott transition \cite{itou2010naturetrilattice,kanoda2006jpsjmitorg}.

Currently, Na$_4$Ir$_3$O$_8$ is one of the most compelling frustrated QSL candidate in three-dimensions (3D) \cite{commentpyrochlore} where Ir$^{4+}$ ions with effective $J_{\rm eff}=\frac{1}2$ form a corner-sharing lattice of triangles named "hyperkagome" \cite{okamoto2007prlna4}. Strong spin-orbit coupling, SOC, has been identified as an important ingredient in the Hamiltonian. Several possible theoretical scenarios have been proposed thus far, including a classical long-ranged order-by-disorder 120$^o$ coplanar ground state \cite{hopkinson2007prlobd} which in the quantum limit melts into a gapped QSL \cite{lawler2008prlcoplanarorder} and a gapless QSL \cite{lawlerprl2008gaplesssl, zhou2008prlgaplesssl}. More generally, iridates appear as an ideal playground for the study of novel physics governed by strong SOC in competition with Coulomb repulsion, crystal field effects, and inter-site hopping. This has led theorists to promote, for example, the Heisenberg-Kitaev model \cite{kimchi2014prbkitaev} and the spin-orbit Mott insulator \cite{jackeli2009MITSOC}. Na$_4$Ir$_3$O$_8$ also appears to be close to an insulator-metal transition \cite{singh2013prbna4,balodhi2014arxivna4-x,fauque2014arxivna3+x,takayama2014scirepna3hyper,chen2013prbMIT} and weak Mottness has been proposed as a possible scenario for a spin-liquid ground state \cite{podolsky2009prlqcpmit,podolsky2011prbsocmit}, like in the case of organic triangular spin-liquids \cite{kanoda2006jpsjmitorg}.

The macroscopic susceptibility, $\chi$, is typical of $J_{\rm eff}=\frac{1}2$ moments with antiferromagnetic interactions $J$ $\sim 300~K$ \cite{okamoto2007prlna4,singh2012prbhtse}. Heat capacity, $C$, shows no sign of a bulk transition and has two remarkable features: at 24~K $\sim J/10$ a broad maximum and at low temperatures $C=\gamma T + \beta T^n$ where 2 $<n<$ 3 \cite{okamoto2007prlna4,singh2013prbna4}. Both $\chi$ and $C$ suggest the presence of a gapless ground state \cite{lawlerprl2008gaplesssl, zhou2008prlgaplesssl}. Magnetocaloric measurements suggest a quantum critical behavior in zero-field \cite{singh2013prbna4}. Although no signs of a bulk transition exist, there is a small splitting of the field-cooled (FC) and zero field-cooled (ZFC) magnetization at $T\sim6$~K \cite{dally2014arxivna4}. Initially, this splitting was proposed to be associated with a $\sim1\%$ defect or impurity term \cite{okamoto2007prlna4,singh2013prbna4}, but recent $\mu$SR measurements \cite{dally2014arxivna4} suggest quasi-static short range spin correlations appear below $T=6$~K.

Probing the role of these defects on the physics of this compound, disentangling their contribution to the susceptibility from that of intrinsic origin and revealing the low energy spin dynamics are the central unexplored focus of the $^{23}$Na and $^{17}$O NMR studies presented in this Letter, down to 1.2~K . After explaining the site assignment and respective advantages of both probes, we are able to study different aspects of the underlying physics. The $^{17}$O shift points to a plateau in the local susceptibility at low temperatures which may be viewed as a partial confirmation of the predictions of \cite{okamoto2007prlna4,singh2012prbhtse}. The evolution of the broadening of $^{17}$O and $^{23}$Na lines reveals a bulk transition to a frozen state at T $\sim$ 7~K with a magnetic moment, $\mu_{\rm Ir } \sim 0.27 \mu_{B}$. Relaxation data, $T_1^{-1}$, on $^{23}$Na line confirms a phase transition at T $=$ 7.5~K while at high temperature, it has very little $T$-variation which serves as an additional signature of the spin liquid regime in this exotic geometry.

The synthesis and the quality assessment of our sample are detailed in \cite{SI}.
NMR measurements on $^{23}$Na, nuclear spin $I=\frac{3}{2}$ and gyromagnetic ratio $^{23}\gamma/2\pi$ = 11.262~MHz/T, and $^{17}$O, $I=\frac{5}{2}$ and $^{17}\gamma/2\pi$ = 5.772~MHz/T, were performed in a pure sample and, for comparison, in a depleted sample (Na$_{4-x}$Ir$_3$O$_8$), in fixed and variable field magnets in a range of 4~T to 11~T using home-built probes. Fourier-transformed spectra were obtained using a standard Hahn echo pulse sequence. The frequencies are given in comparison to a $^{23}$Na and $^{17}$O reference, NaCl solution, used as the zero frequency. Spin-lattice relaxation rates were obtained with a saturation recovery pulse sequence.

\begin{figure}
\centering
\includegraphics[width=0.8\linewidth,height=0.4\textheight]{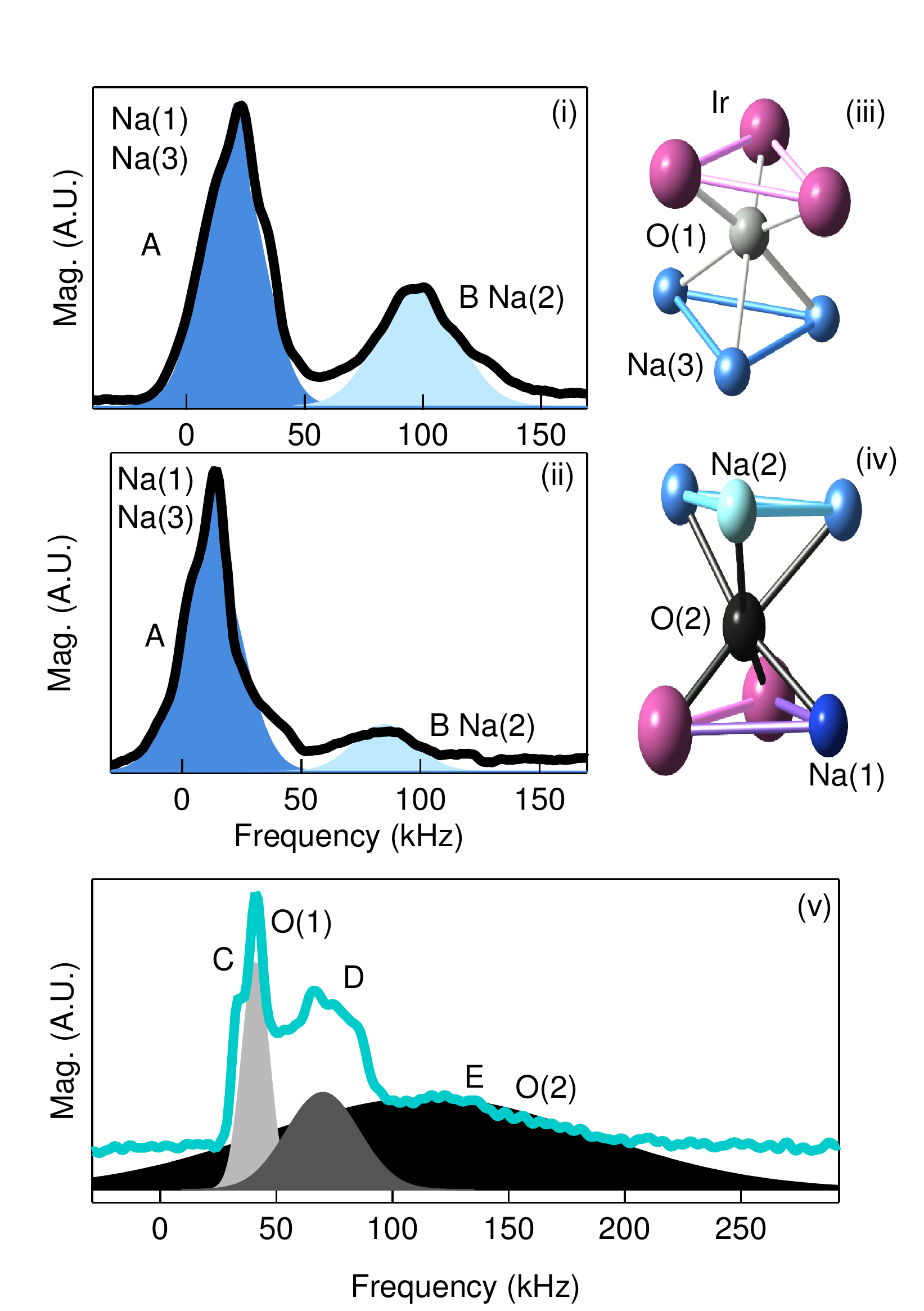}
\caption{Spectra taken at 300~K with a frequency of 68MHz for $^{23}$Na in Na$_4$Ir$_3$O$_8$ (i) and Na$_{4-x}$Ir$_3$O$_8$ (ii) and with a frequency of 43.3MHz for $^{17}$O in Na$_4$Ir$_3$O$_8$ (v). Solid lines represent raw data; Gaussian curves indicate the location of the different sites: peak A as blue, peak B as pale blue, peak C as light gray, peak D as dark grey, peak E as black. The local environment of (iii) O(1) and (iv) O(2) are shown. \label{spectra}}
\end{figure}

The crystal structure has three $^{23}$Na sites. Na(1) combined with three Ir form the tetrahedron of the pyrochlore lattice; Na(2) and Na(3) also form a network of corner-shared tetrahedra, each having a 75\% occupancy \cite{okamoto2007prlna4}. In the unit cell, the ratio of Na(1/2/3) is calculated as (4:3:9). The spectrum of the central transition of Na$_4$Ir$_3$O$_8$
is shown in Fig.~\ref{spectra}(i). While the three expected sites could not be resolved, the spectrum has two distinct peaks, labeled A and B, with weights of  75(6)\% and 25(6)\% when corrected for the relaxation rates, respectively. Since Na(3) is 56\% of the total Na, part of peak A must be associated with Na(3). A further analysis of the first order satellite lines \cite{SI} supports this conclusion. Comparing the spectrum for Na$_{4-x}$Ir$_3$O$_8$ to Na$_4$Ir$_3$O$_8$ (Fig.~\ref{spectra}(ii) to (i), respectively), peak B has lost a considerable amount of intensity. Since the occupancy of Na(2) and Na(3) decreases with increasing $x$ in Na$_{4-x}$Ir$_3$O$_8$ \cite{dally2014arxivna4}, peak B is most likely associated with the Na(2) crystallographic site. The last site, Na(1), was impossible to resolve but is expected to account for the rest of the intensity of peak A.

\begin{figure}
\centering
\includegraphics[width=0.95\linewidth,height=0.22\textheight]{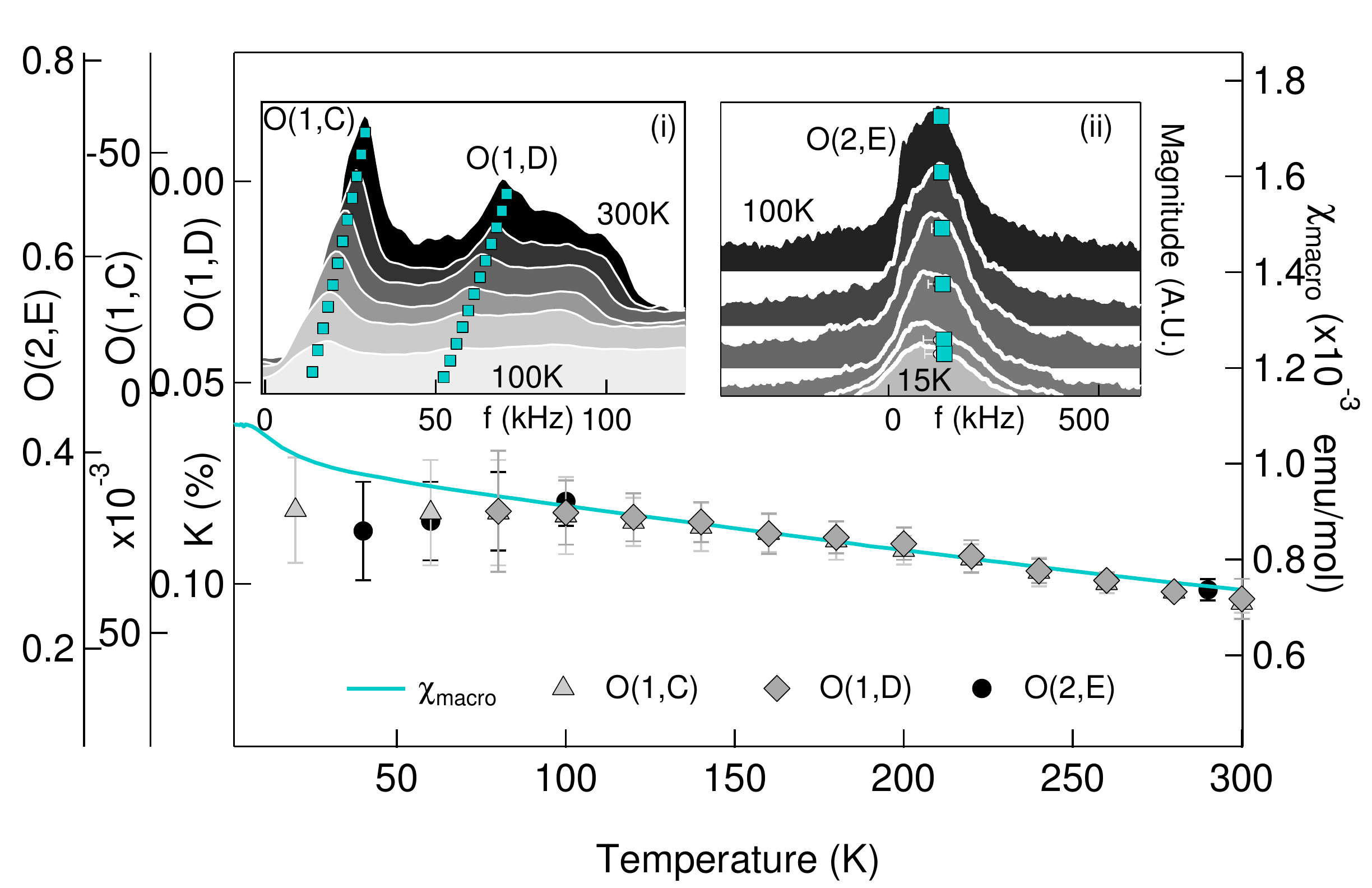}
\caption{The shift of the three $^{17}$O sites (markers, left axis) and the bulk susceptibility (solid line, right axis) as a function of temperature. Inset: (i) $^{17}$O spectra of peaks C and D taken with a frequency of 68.077~MHz from 300~K to 100~K in 40~K intervals. (ii) Isolated spectrum of peak E obtained by contrast with a frequency of 37.518~MHz from 100~K to 20~K in 20~K intervals and 15~K. The blue markers show the bulk susceptibility.  \label{knightshift}}
\end{figure}

$^{17}$O has two crystallographic sites, O(1/2), with ratios 1:3 whose local environments are shown in Fig.~\ref{spectra}(iii) and (iv), respectively. The spectrum
is shown in Fig.~\ref{spectra} with three peaks labeled C, D, E. Since Na(2/3) have a 75\% partial occupancy \cite{okamoto2007prlna4}, the O(1) site has two dominant local environments which, assuming a random binomial occupation, should each account for 10.5\% of the total spectral weight: one fully occupied with three-fold rotational symmetry and one with two of three Na sites occupied with no rotational symmetry. An analysis of the first order satellites \cite{SI} shows that peak C is the only site with high symmetry and therefore corresponds with O(1). Since peak D has a similar spectral weight to peak C (17(5)\% and 15(3)\% respectively), it also corresponds with O(1). The remaining peak E which has the bulk of the spectral weight (67(8)\%) must correspond with O(2).

\begin{figure}
\centering
\includegraphics[width=0.9\linewidth,height=0.28\textheight]{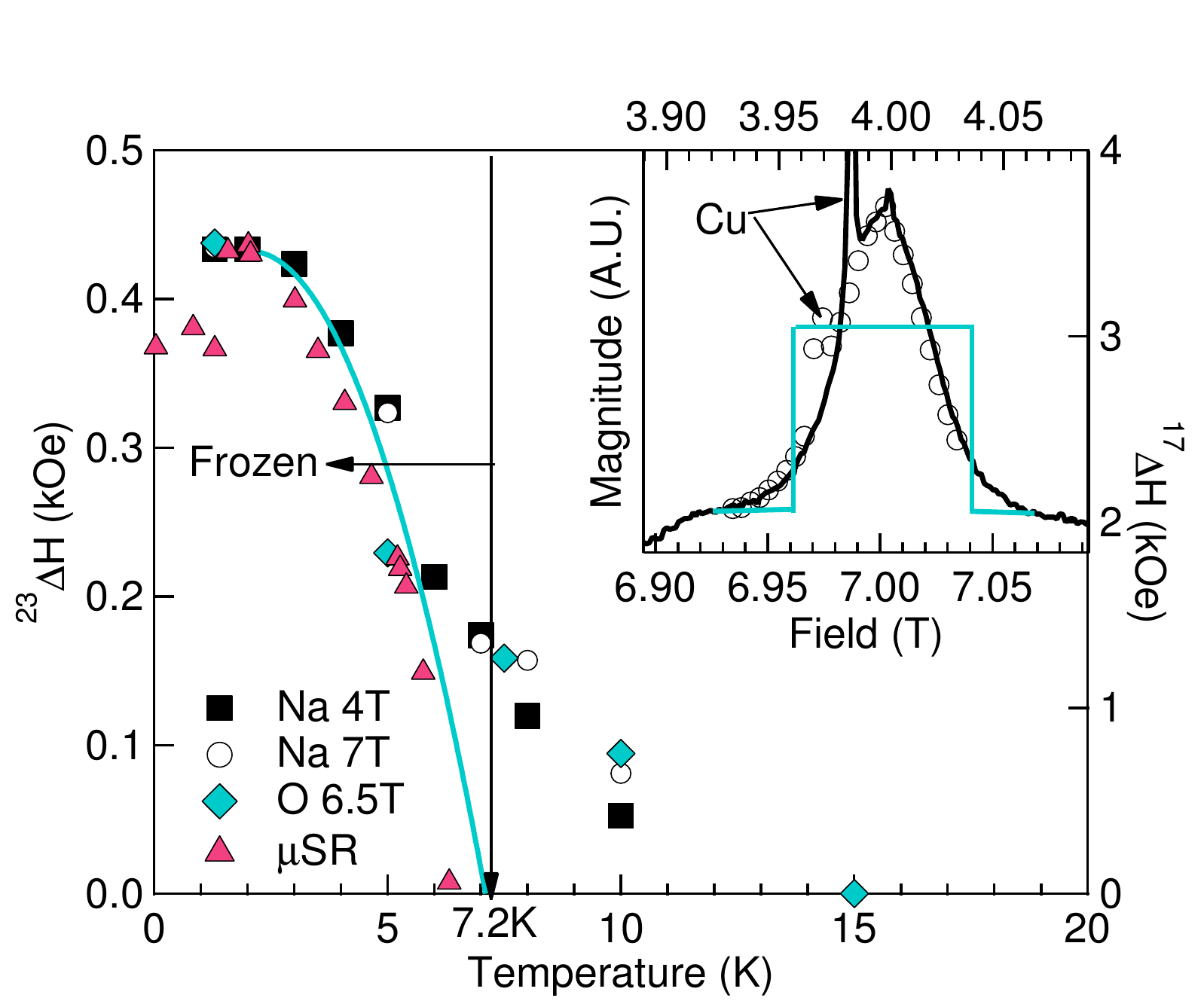}
\caption{FWHM of the Gaussian broadening with respect to 15~K in kOe for $^{23}$Na at 4~T and 7~T (left axis) and $^{17}$O at 7~T (right axis) as a function of temperature below 16K. The $\mu$SR data from \cite{dally2014arxivna4}, normalized to the NMR 1K $\Delta H$, as pink triangles. Inset: $^{23}$Na spectrum at 45.046~MHz for 4~T (solid line) and 78.937~MHz for 7~T (empty circles) at $T=1.3$~K. The spectrum centered at 4~T has been shifted horizontally by 3.005~T and renormalized for comparison. Background Cu is from the experimental setup. The blue line indicates the expected shape of the spectrum if long ranged order were present with the same value of the moment. \label{orderparameter}}
\end{figure}

Most of our NMR static data reported in this Letter have been taken on O sites which are better coupled to Ir moments. We used a combination of the data from the O(C,D,E) lines in order to analyze the shift and the linewidth, our selection dictated by resolution and coupling considerations. The temperature evolution of the spectra are an example (Fig.~\ref{knightshift} insets (i) and (ii)). The O(1,C) peak is resolved the best and can be followed down to 20~K. The O(2,E) peak can be isolated by a contrast experiment using fast repetition rates. Its shift can only be reliably extracted down to 40~K due to substantial broadening.

In the paramagnetic regime, the spectral shift is linearly related to the local susceptibility, $\chi_{\rm loc}$,  by $K(T) = K_0 + A_{\rm hf}\chi_{\rm loc}(T)$ where $K_0$ is a temperature independent orbital contribution and $A_{\rm hf}$ is the hyperfine coupling constant \cite{SI}. The spectral shift tracks the intrinsic susceptibility while the magnetic broadening of the line reflects the distribution of local susceptibilities, in general due to a small amount of defects. Thus, spectral measurements can uniquely disentangle these two contributions to the macroscopic susceptibility. We find a good agreement between $K$ for O(1,C), O(1,D) and O(2,E) peaks and $\chi_{\rm macro}$ between 300 and 100~K and establish that the intrinsic susceptibility levels off below 80~K. The lower $T$-limit of 20~K of our data conservatively indicates the absence of a gap larger than $J/15$.

Looking closer at Fig.~\ref{knightshift}(ii), the linewidth, $\Delta$H, of the O(2,E) peak gradually increases down to 15~K as expected from the Curie tail of $\chi_{\rm macro}$. Below 10~K, $\Delta$H increases significantly more, as seen in Fig.~\ref{orderparameter}. This broadening is best tracked by the $^{23}$Na(A) line which remains well resolved, has a small high-$T$ width and can be studied at various fields due to the Na NMR sensitivity \cite{SI}. Below 7.5~K where the ZFC-FC susceptibility measurements split, $^{23}\Delta$H is the same for 4~T and 7~T which is a landmark of static freezing. This frozen phase is fully settled by 2K below which the linewidth is constant, consistent with $\mu$SR measurements \cite{dally2014arxivna4}. If this broadening were due to the onset of an ordered phase, the well-defined internal fields would cause the powder spectrum to be rectangular. 
As shown in the inset of Fig.~ \ref{orderparameter}, this is not the case and the identical spectra at 1.3~K can be approximately fitted by a Gaussian indicating a disordered freezing of Ir moments. The absence of any residual narrow spectral contribution which would be typical of a paramagnetic phase indicates a freezing of {\it all} moments.

In the frozen phase, the FWHM of the Gaussian broadening with respect to 15~K tracks the order parameter as in the muon data \cite{dally2014arxivna4}. Since the  hyperfine constant 26(5)~kOe/$\mu_{B}$ is most accurately determined for the O(2,E) peak from the $K-\chi$ plot in the $T > 100$~K paramagnetic regime (see Fig.~\ref{knightshift}), the magnetic moment of Ir$^{4+}$ can be best extracted from $^{17,E}\Delta H$ at 1.3~K . In the framework of a Gaussian lineshape, $\mu_{\rm Ir} = \frac{2\ln{2} \Delta H z^{1/2}}{A_{\rm hf}}$ where $z$ is the number of Ir coupled to the O. This yields 0.27(4)$\mu_{B}$. This estimate is comparable to the observed $\mu_{\rm Ir}$ from neutron measurements in other iridates such as Sr$_2$IrO$_4$ \cite{lovesey2012jpcmsrir2o4irmoment, yefeng2013prbsr2iro4moment} and Na$_2$IrO$_3$ \cite{ye2012prbna2iro3moment}.

We now turn to the spin dynamics as revealed by our relaxation data. We have measured $T_1^{-1}$ on the $^{23}$Na(B) peak which is well-defined until below 15~K. It has sufficient coupling to the Ir site with better separation than the O sites which allows us to isolate the relaxation components. (23)
For the nuclear spin of $^{23}$Na, \textit{I}=$\frac{3}2$, the magnetization recovery is described by only two components \cite{narath1967physrevSLR}
\begin{equation}
M(t)=M_0[1-(a e^{-(6t/T_1)^{\beta}}+(1-a) e^{-(t/T_1)^{\beta}})]
\end{equation}
where M$_0$ is the equilibrium magnetization, \textit{a} measures the weight of the short and long components of $T_1$ and is typically $\frac{9}{10}$. $\beta\neq 1$ is introduced here to account for the distribution of relaxation rates \cite{johnston2006PRBbetaexp} which is related to the degree of disorder in the system.

\begin{figure}
\centering
\includegraphics[width=\linewidth,height=0.3\textheight]{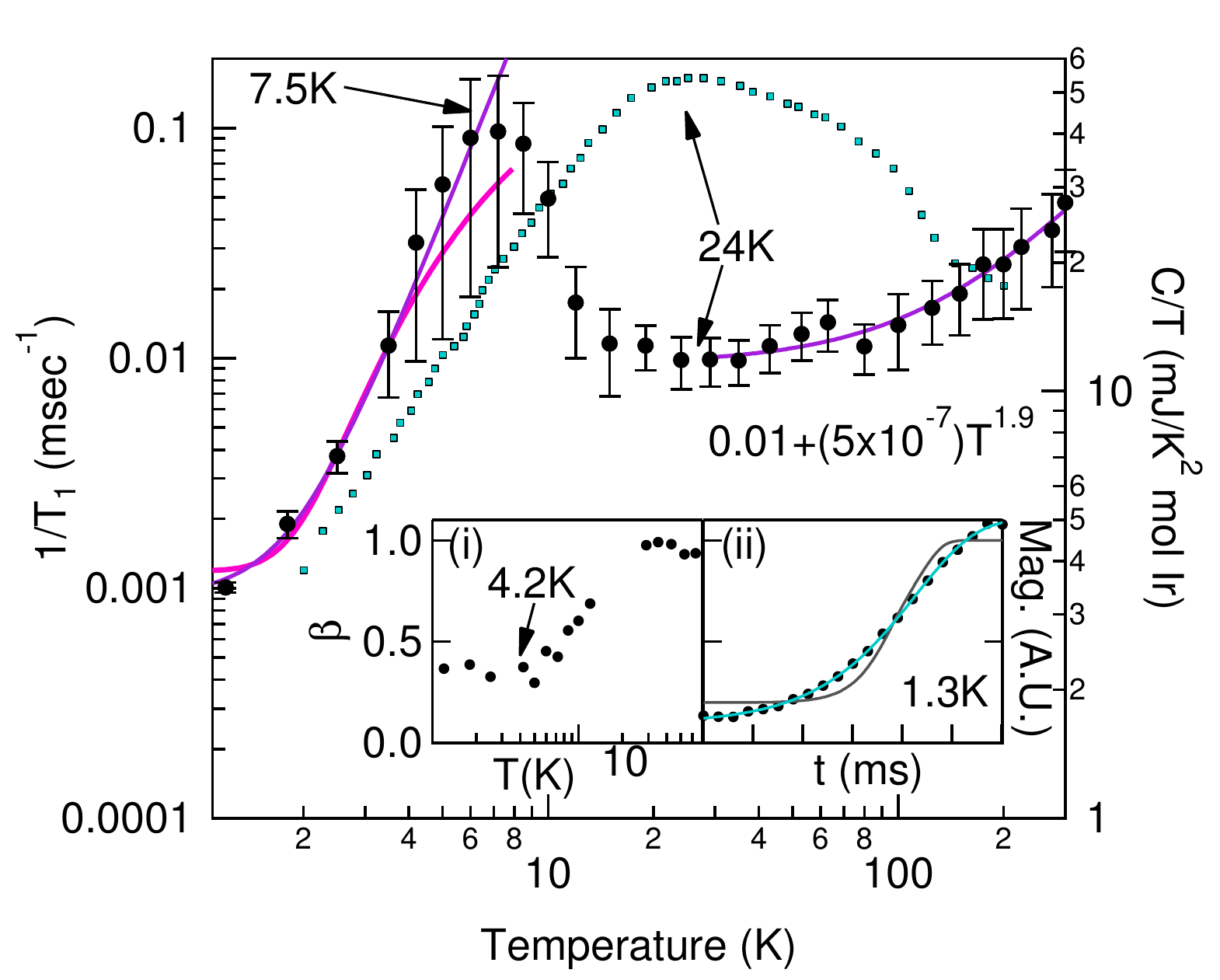}
\caption{$^{23}T_1^{-1}$ (left axis, black circles) and C/T from \cite{okamoto2007prlna4} (right axis, blue squares) as a function of temperature. The purple lines are fits. Inset: (i) The stretch exponent, $\beta$, as described in the text as a function of temperature. (ii) The relaxation curve (solid circles) at 1.3~K fit with a stretched exponent (blue line, $\beta=0.4$) and exponential fit (grey line, $\beta=1$). \label{T1}}
\end{figure}

$T_1^{-1}$ decreases very smoothly until $T\sim 24$~K, with a phenomenological fit $T_1^{-1} = A + B T^{\alpha}$ (Fig. \ref{T1}) where the constant term $A=0.010(2)$~ms$^{-1}$ dominates. The 300~K value is consistent with  the exchange narrowing limit of Heisenberg antiferromagnets given by Moriya's theory~\cite{moriya1956progtheorphys}, $1/T_1\sim A_{\rm hf}^2 /J = 0.018(9)$~ms$^{-1}$. Given the only slight decrease of $1/T_1$, $1/T_1 T$ increases when $T$ decreases which could indicate a moderate strengthening of correlations until reaching a maximum at the phase transition at $T\sim7.5$~K. Below 7.5~K, $T_1^{-1}$ drops by two orders of magnitude which is typical of the critical slowing down of spin fluctuations when entering a frozen phase. The relaxation rates become distributed below 30~K where $\beta$ begins to decrease monotonically (Fig.~ \ref{T1}(i)).
For a gapped state, $T_1^{-1} \sim \exp{(-\Delta/T)}$ at low temperatures. An exponential ($\Delta \sim 12(2)$~K) and a phenomenological power law (power $\sim 3.8(3)$) fit were made for $T<$7~K. There is no substantial difference between these two fits. Both fits require a vertical offset of $\sim 0.001$ which might originate from a minute amount of paramagnetic uncoupled moments. We note that the absence of gapped behavior in the specific heat might favor the power law for $1/T_1$. Quite surprising is the  discrepancy between our $1/T_{1,{\rm NMR}}$ spin lattice relaxation rate and the $\mu$SR relaxation rate $1/T_{1,\mu {\rm SR}}$ \cite{dally2014arxivna4}. $1/T_{1,\mu {\rm SR}}$ keeps {\it constant} which was argued to be the signature of configurationally degenerate phases with {\it fluctuating} order \cite{dally2014arxivna4}. Recent LDA calculations on model systems \cite{foronda2015prlmuondft,bernardini2013prbmuon} show the $\mu^{+}$, as a positive charge, might induce local deformations in its environment which could affect the dynamics of Ir$^{4+}$ close to the muon probe.

From the experimental results  to be discussed below, two $T$-regions emerge from our study: one typical of the ideal hyperkagome, Na$_4$Ir$_3$O$_8$, at high $T$ and one whose physics deviates from the ideal Hamiltonian(s) proposed, leading to a fully frozen phase below $\sim 7$~K.

Based on our NMR data we draw the conclusion that a {\it bulk} and {\it static} disordered spin freezing occurs at $T_f$ with {\it inhomogeneous} dynamics in the ground state. Bulk spin freezing might appear to conflict with the signature of a ZFC-FC splitting in $\chi_{macro}$ which was attributed to "defect" spins. This could have been interpreted as a marginal spin glass transition implying a weak coupling of such defects, e.g.  mediated through a spin liquid background but these defect spins, which dominate the low-$T$ susceptibility, might act simply as a fingerprint of the bulk physics.

In view of the 75\% partial occupancy of Na \cite{okamoto2007prlna4}, we propose that some local charge disorder occurs. This certainly affects the Ir$^{4+}$ environments and leads to a distribution of magnetic interactions on the hyperkagome lattice, especially if interactions are dominantly due to direct exchange as argued in many papers \cite{chen2008prbspinorbna4}. Deviations to equilateral interaction  triangles have always led to transitions at $T\ll J$ such as observed in Volborthite \cite{yoshida2011prbvolborthite} and Vesignieite \cite{quilliam2011prbvesignieite,Yoshida2013JPSJvesignieite} kagome-based compounds. Digging out why a distribution of interactions could lead to such a weak or no signature in $\chi$ and $C/T$ respectively should be addressed theoretically which may help discriminate between models proposed for this iridate. We note that in all investigated models, the $T^2$ and possibly $T$-linear ultimate behavior of the specific heat has played a central role. This should be now toned down in view of the freezing evident in our data which invalidates the use of $T<T_f$ thermodynamic quantities as characteristic of the spin liquid behavior.

Above the freezing temperature, a broad range of temperatures probes the effects of frustration on the spin liquid state since $J/2T_f \sim 20$. First, in the $T_f$-$2\;T_f$ range, the increase of $1/T_1$ may be interpreted as a slowing down of magnetic fluctuations. Whether this extended range of critical fluctuations could be explained by disorder or it rather appears as a signature of a crossover region coinciding with the maximum of the specific heat at $T\sim3\;T_f$ remains speculative. Above $3\;T_f$, three landmarks of the spin liquid regime now clearly appear, the pseudo-gap like behavior of $T_1^{-1}$, the leveling off of $\chi$ at 80~K confirmed by our shift measurements, which both add up to the broad maximum in $C/T$ at 24~K.  Various models have been explored which we discuss with respect to our results:

(i) A fermionic approach naturally leading to a spinon Fermi surface. The maximum of $C/T$, far too high for a spin glass freezing \cite{martin1980PRBcspinglass}, could be the landmark of a crossover from a $U(1)$ spin liquid to a $\mathbb{Z}_2$ one with a paired spinon state and line nodes in the gap below 20~K \cite{zhou2008prlgaplesssl}. The mixing with triplet states induced by SOC or Dzyaloshinkii-Moriya interactions could explain why the susceptibility keeps its Pauli-like behavior. Yet, our $1/T_1$ data, if intrinsic, contradicts the existence of a gap as it does not decrease,
below the maximum of the specific heat. Furthermore, for a Dirac U(1) spin liquid, one expects $1/T_1\sim T^{\eta}$ where $\eta$ is related to the shape of the correlation function and remains unknown~\cite{ran2007prlheisenbergmodelkagome}. This is not what is observed here since $T_1$ is dominated by a constant term for $T>20$~K.

(ii) The transition to metallicity, observed either under pressure \cite{takagiunpublished} or in depleted Na samples \cite{singh2013prbna4,balodhi2014arxivna4-x,fauque2014arxivna3+x,takayama2014scirepna3hyper}, might indicate the proximity to a quantum critical point. In this context, modeling has focused on the metallic rather than insulating side \cite{podolsky2009prlqcpmit,podolsky2011prbsocmit}; here, the $T-$dependence of $T_1^{-1}$ is a crucial test.

(iii) A 72-sites valence bond crystal has also been proposed as a trial ground state \cite{bergholtz2010prlsymmetry} but, with a gap of the order of $J$, is not relevant. However, a valence bond glass state with a transition at $\sim J/10$ as argued for the kagome lattice is worth further exploration \cite{singh2010prlvalencebondglass}. The low-$T$ behavior of $C(T)$ might originate from the free energy landscape typical of disordered systems such as conventional spin glasses \cite{martin1980PRBcspinglass}.

In conclusion, our $T_1$ data opens up the space of physical quantities to be discussed in future advanced modelings altogether with the constant susceptibility and  the maximum of the specific heat which have been a central issue in the to-date discussions. Whether a slight disorder in the interactions might bear an explanation for the disordered static freezing with no signature in the specific heat is also a major avenue for future work on the hyperkagome lattice.

\begin{acknowledgments}
The authors would like to thank C. Payen for help with characterization and S.-H. Baek, B. Fauqu\'{e}, Y. Singh, and L. Balents for discussions. Y.O. thanks Z. Hiroi for helpful discussion and for his support on sample preparation performed at ISSP, University of Tokyo. This work was supported by the French Agence Nationale de la Recherche under Grant "SPINLIQ" No. ANR-12-BS04-0021, by Universit\'{e} Paris-Sud Grant MRM PMP, and by JSPS KAKENHI Grant No. 25800188.
\end{acknowledgments}

\bibliography{shockleybibliography}

\end{document}